# Polarized emission of GaN/AlN quantum dots :
# single dot spectroscopy and symmetry-based theory

R. Bardoux[1,2], T. Guillet[1,2], B. Gil[1,2], P. Lefebvre[1,2], T. Bretagnon[1,2], T. Taliercio[1,2], S. Rousset[1,2], and F. Semond[3]

  1 : Université Montpellier 2 – Groupe d'Etude des Semiconducteurs (GES).
    CC 074. F-34095 Montpellier Cedex 5, France.
  2 : CNRS – UMR 5650.
  3 : CNRS – Centre de Recherche sur l'Hétéro-Epitaxie et ses Applications. Rue Bernard Grégory. F-06560 Valbonne, France.

## Abstract

We report micro-photoluminescence studies of single GaN/AlN quantum dots grown along the (0001) crystal axis by molecular beam epitaxy on Si(111) substrates. The emission lines exhibit a linear polarization along the growth plane, but with varying magnitudes of the polarization degree and with principal polarization axes that do not necessarily correspond to crystallographic directions. Moreover, we could not observe any splitting of polarized emission lines, at least within the spectral resolution of our setup (1 meV). We propose a model based on the joint effects of electron-hole exchange interaction and in-plane anisotropy of strain and/or quantum dot shape, in order to explain the quantitative differences between our observations and those previously reported on, e.g. CdTe- or InAs-based quantum dots.

**I. INTRODUCTION**

Quantum dots (QDs) based on a variety of semiconductor compounds have attracted considerable interest in the past few years, not only because of potential applications to improved solid-state lasers or to controlled single photon emitters, but also because of some original physical properties. Among these properties are those which can be manifested by spatially-resolved photoluminescence (the so-called micro-PL), which permits to analyze light emission from a single QD. Single-dot PL spectra are usually composed of several discrete, narrow lines that result from the recombination of different, neutral or charged, excitonic complexes. Polarization sensitive studies of the fine structure of such single QDs have been reported for several types of self-organized QD systems: CdSe [1], InP [2], CdTe [3], InAs [4-7]. In these material systems, the neutral exciton spectrum of single QDs exhibits a doublet of lines that are linearly polarized along two perpendicular directions. These lines are generally split by an energy that arises from the joint effects of the long-range electron-hole exchange interaction and of some in-plane anisotropy. The latter can result either from the elongation of the QD shape or from the anisotropy of the strain or even the strain-induced piezoelectric potential [6]. The short-range contribution to exchange interaction rather splits the radiative excitonic doublet from the non-radiative one. This can be evidenced by spectroscopic experiments under magnetic field [3].

It is worth noting that the intensities of the cross-polarized components of the single-dot PL are quite generally comparable. We say that the polarization degree is small in such cases, contrary to the situation encountered for strongly anisotropic zero-dimensional systems. For instance, strongly elongated interface defects in GaAs [8] or CdTe [9] quantum wells, induce a strong polarization degree of the radiative doublet. The latter is still split by the exchange interaction, but it is the light-hole-to-heavy-hole valence-band mixing that modulates the oscillator strengths of the different components, in case of anisotropic confinement. This has been modeled, for instance, for the case of quantum wires [10]. The complexity of the valence band mixing has recently been evoked as a possible explanation of the observed polarization features exhibited [7] by some InAs QDs: although a thorough (albeit single-valence-band) modeling of multi-exciton recombination, including multiply charged complexes, provides satisfactory assignment of the observed PL lines and of their polarization properties, the model fails at explaining why the polarization does not follow systematically crystallographic axes.

Newcomers in the field of single-QD spectroscopy are those based on wide band-gap, wurtzite, group-III nitrides such as GaN/AlN QDs. The wurtzite symmetry generally induces internal electric fields of several MV/cm along the (0001) growth axis. This results in giant quantum-confined Stark effect that, when the QD height is increased over ~2-3 nm, red-shifts exciton energies over hundreds of meV [11,12], yields non-conventional recombination dynamics [13], increasing radiative lifetimes over several orders of magnitude [14] and induces strong electron-hole dipoles that are especially sensitive to their electrostatic environment [15].

In spite of these challenging experimental conditions, single-dot spectroscopy has been performed on wurtzite GaN/AlN QDs either grown along the (0001) direction [15-19] or along the non-polar (11-20) direction [20,21], which reduces electric field effects. Biexcitonic recombination was identified and studied [16,19,22], as well as spectral diffusion effects [15], Stark shift [23,24], and controlled single photon emission has proven to be at hand [17]. However the knowledge of the excitonic states in nitride QDs is still incomplete : especially electrons and holes experience an additional lateral confinement due to the strain distribution inside the QD through piezo-electric fields [25], which is inaccessible experimentally; valence-band mixing is important due to the small A-B valence band splitting; direct and exchange Coulomb interactions are strong due to the small excitonic Bohr radius, comparable to the QD lateral size. A theoretical model including all those effects, with very similar orders of magnitude, on the same footing, is still lacking.

Polarization sensitive studies of single nitride QDs have however only been proposed, so far, on (In,Ga)N/GaN QDs [26]. Strong linear polarization has been observed in this case, with two preferential directions following the (11-20) and (-1100) crystallographic axes, while such a polarization is not authorized by the crystal symmetry in the bulk material. The authors of Ref. 26 assign the polarization properties to the in-plane elongation of their QDs that would result from the specific mechanism of QD formation (spinodal decomposition).

In this paper, we present a detailed study of linearly polarized micro-PL taken from GaN/AlN QDs grown by molecular beam epitaxy in the Stranski-Krastanov growth mode, on Si (111) substrates. We evidence a strong linear polarization degree of the QD transitions, which is the signature of the valence band mixing induced by the in-plane anisotropy of strain and/or quantum dot shape. We then propose a multi-band excitonic model based on the joint effects of electron-hole exchange interaction and in-plane anisotropy, which treats the confinement within a rough and symmetry-based approximation. We emphasize the qualitative and quantitative differences between nitride QDs and zinc-blende QDs, and propose an interpretation of our experimental results.

## II. EXPERIMENTAL RESULTS

### II.1. Experiment

We present single QD spectroscopy performed on GaN/AlN self-assembled QDs. The sample consists in a single plane of dots grown via the Stranski-Krastanov mode by molecular beam epitaxy [27] along the (0001) axis. The dots lie on a 400 nm AlN epilayer, itself deposited directly on the (111) face of a Si substrate [28]. A 175 nm AlN cap layer covers the QDs. By stopping the rotation of the sample holder during growth, a gradient of QD density is produced. In the most dilute parts of the sample, the density can therefore be made as low as $10^9$ cm$^{-2}$. Moreover, the QD height has been chosen small enough to minimize the effects of internal electric fields, thus yielding photoluminescence (PL) emission energies close to the band gap of GaN biaxially compressed onto AlN. Previous µPL studies on the same sample have evidenced a laser power-dependent spectral diffusion of the QD transitions arising from some QDs, with an amplitude of a few meV [15]. In the present work, as it will be discussed later, we focus exclusively on QDs which do not exhibit such a spectral diffusion.

The sample is cooled to T=10K in a continuous flow helium cryostat dedicated to µPL studies. The photoluminescence is excited by the second harmonics of a continuous wave argon laser ($\lambda$=244 nm, h$\nu$=5.07 eV) which is focused through a 36x Cassegrain microscope objective. The laser energy is below the AlN bandgap, so that electron-hole pairs are created directly within the QDs and their wetting layer. The diameter of the excitation spot is ~1.5 µm, and the excitation power density is about 200 W/cm$^2$ in the focal plane of the microscope objective. The linear polarization of the emitted light is analyzed with a calcite Glan polarizer. A 60 cm spectrometer with a 1200 grooves/mm grating, coupled to a nitrogen-cooled CCD camera, is used for detection. A spectral resolution of 1 meV is reached. Prior to analysis through the spectrometer, the PL light is converted into circular polarization with a λ/4 waveplate in order to set free of the spectrometer polarization response function.

**II.2 Polarized photoluminescence spectrum of a single quantum dot**

We have studied the polarization properties of the PL of single QDs. We collect the photons emitted perpendicular to the sample surface. Isolated emission lines, with linewidths limited by our spectral resolution, have been followed as a function of the angle of the analyzer over 180°. A typical result is presented in Fig. 1.a. The PL spectrum shows five sharp peaks which correspond to different QDs lying within the excitation spot, as well as smaller peaks and a structured background which are attributed either to neighboring QDs or to the effect of the electrostatic fluctuations in the environment, i.e. spectral diffusion [15]. The angular dependence of the intensity of 3 lines is reported on the angular chart in Fig. 1.b. As expected, this dependence is well accounted for by the equation:

$$I(\theta) = a + b\cos^2(\theta - \theta_0) \qquad (1)$$

We generally observe a strong degree of linear polarization of these transitions. However we may notice three important points : (i) the lines are not fully polarized ($a \neq 0$); (ii) their angle of polarization $\theta_0$, determined with an uncertainty of 10°, differs from line to line, and it does not follow specific crystallographic orientations; (iii) within our spectral resolution of 1 meV, we do not observe doublets of distinct lines with orthogonal polarizations and separated by less than 10 meV, as usually observed for InAs or CdTe quantum dots [1,3,4].

We may here emphasize the experimental difficulties related to the existence of a strong spectral diffusion. In Fig. 1.a, the intensity of the peaks at 3.59 and 3.63 eV, which is not represented in Fig. 1.b, does not follow Eq. (1) despite strong intensity variations among the different spectra. These variations are therefore attributed to the fluctuating electrostatic environment of the corresponding QDs, which might modify their capture rate. This experimental constraint obliged us to establish a specific protocol in order to unambiguously measure the polarization properties of the emission : during the recording of PL spectra, the sequence of angles was chosen arbitrarily, and not monotonous. Only the peaks for which the intensity properly follows Eq. (1) were selected for further interpretations. Moreover the fluctuations in the environment of the QD do not only influence the intensity of the lines, but also their energy. Even if small shifts of ~1 meV have been occasionally observed between spectra recorded for various angles, we never observed any clear correlation of these shifts with the angle of the polarizer, as previously reported for a doublet of two split and cross-polarized lines [1,3,4]. We therefore assign the small observed shifts to the spectral diffusion only.

### II.3. Statistics of the polarization properties

Similar measurements have been performed for about 50 QDs. The measured polarization degree $P$ strongly differs from line to line. It is defined as :

$$P = \frac{I_{max} - I_{min}}{I_{max} + I_{min}}. \qquad (2)$$

$P$ can reach up to 90%, as shown for the well isolated emission line presented in Fig. 2. Twenty of the studied QDs exhibit an unambiguous linear polarization, whereas the polarization degree of the remaining QDs, smaller than 30%, was difficult to measure accurately because of spectral diffusion effects. The statistical distribution of $P$, presented in Fig. 3.a, seems rather uniform between 0 and 100%. The measured polarization angles (Fig. 3.b) are also continuously distributed. The direction close to 0° is more represented than the others but the angles are not fully correlated to the crystallographic axes of the substrate or of the underlying AlN epilayer.

### II.4. Discussion

These experimental results show that the PL of a single QD, apart from spectral diffusion effects, mainly presents a single transition line, which is often strongly linearly polarized. This seems contradictory with the two cross-polarized bright states that have been observed in other QD systems. As it will be described in details in the next section, our model yields a spectrum for a single QD consisting in three optically active states (in the lowest band), which are fully linearly polarized along x, y and z directions, and may be degenerate or lie within a few meV. In our experiments, the PL signal never vanishes completely at a given in-plane polarization ($a \neq 0$ in Eq. (1)). Our interpretation therefore is that the above-mentioned x and y-polarized PL lines correspond in fact two quasi-degenerate states (within the limit of our spectral resolution of 1 meV), and that their intensities differ in the two polarizations. Let us note that in the case of a large splitting (larger than 3 meV) and a thermal equilibrium within the doublet, the higher-energy state would hardly be distinguished from the background signal, but the lower state should be *fully* linearly polarized ($a=0$), which is not the case in our experimental spectra.

Concerning the thermalization of the carriers, we have checked that the measured degree of polarization remains the same when the linear polarization of the exciting laser is changed, showing that the spin of the carriers is random after relaxation towards the confined states: all states are equally populated. We can therefore conclude that the PL degree of polarization probes the degree of polarization of the oscillator strengths of the different states involved. Its large value (up to 90 %) and the absence of observable splitting between the cross-polarized transition lines, is in striking contrast with the polarization properties of excitonic spectra on InAs or CdTe QDs, for which two split and cross-polarized lines with similar intensities are observed. These two features will be analyzed within the framework of the model developed in the following section.

The nature of the emitting states is an important issue. Although our heterostructure sample is not intentionally doped, we cannot ascertain that the PL arises from neutral excitons: the QDs may be charged due to the photo-excitation or the vicinity of donors and acceptors, and therefore trions or other types of charged excitonic complexes could be observed. On the other hand, we can safely discard emission from neutral many–exciton states, like e.g. biexcitons, because the intensities of all the investigated transitions exhibit a linear or slightly sub-linear dependence with the excitation power density.

### III. MODEL

The problem of the polarization properties and the fine structure of the QD emission is twofold : (i) the single-carrier (electron or hole) eigenstates of the QD have to be precisely known in terms of their momentum, spin and envelope wavefunctions; (ii) then the few-body wavefunctions have to be properly built. The following model is intended to carefully describe the momentum and spin wavefunctions of the ground confined states of electrons and holes in the QDs, and to account only qualitatively for the confinement. Moreover, the model is restricted to neutral excitons and negative trions, since our experimental data do not call for more complicated multi-particle systems. The rigorous derivation of states with more particles requires a much more cumbersome model [29,30].

Compared to recent theoretical works, our model intentionally keeps the simplicity of a matrix description of the single-particle states in order to clearly exhibit the main driving parameters and to allow for a comparison with QDs based on other materials. A more precise treatment of the confinement can be obtained through a **k.p** model [25,26,31], a tight-binding model [32] or a pseudopotential method [33]. Full numerical resolution, in these models, would provide the ground as well as excited states for electrons and holes, but it would somehow mask the driving parameters. Moreover such models require many inputs on the exact shape, composition, strain and electric fields in the QDs, not to mention band offsets and valence band parameters, all of which retain some uncertainty up to now for GaN QDs. Here we are interested in the polarization properties and fine structure splittings rather than the absolute values of the full set of recombination energies. Concerning the many-particle states, it is challenging to go beyond neutral and singly charged excitons : the full second quantization treatment including exchange terms and anisotropy has only been solved for InAs quantum dots within the approximation of a single heavy-hole valence band [7], but such an approximation can absolutely not be applied in the present work, given the complexity of valence band mixing schemes in nitrides where, e.g., the A and B bands are only separated by about 10 meV.

In the following we will first focus on the single-carrier properties, especially the symmetry of hole states in the valence band. We will then derive the energies and polarization-dependent oscillator strengths of neutral excitons and trions.

**III.1 Single-carrier eigenstates**

The basic elements of this section have been developed by K. Cho in his general paper dedicated to symmetry breaking effects in zinc blende and wurtzite semiconductors [34]. The important point of this paper that we wish to outline here is that kinetic energy terms, which are a bilinear form of the components $k_i$ of the wave vector **k**, and components $\varepsilon_{ij}$ of the strain tensor $\bar{\bar{\varepsilon}}$, can be treated on the same footing, since appropriate combinations of them transform according to the same irreducible representations of the $C_{6v}$ point group symmetry. Table 1 summarizes these relationships. Those are the basic building blocks that are necessary to go further in the quantum description of excitons, trions and biexciton states that are all built from conduction (electron) and valence (hole) band states.

For QDs, the wave vector **k** is not a good quantum number but we will keep a description in terms of **k**, which scales as the inverse of the QD dimensions. Let us underline that we deliberately treat the confinement of the carriers through a very rough approximation, which properly introduces all confinement-induced band-mixing terms with their correct symmetries, even if it provides a too simple estimation of the confinement energy. Therefore the absolute value of the calculated spectra should be considered as indicative, but the energy splittings and polarization properties are qualitatively correct even if their value may be slightly different from our calculation. As stated in the introduction, our purpose is to provide an analytic model with the minimum number of parameters and with the correct symmetries needed to investigate the polarization properties of exciton complexes in QDs. This allows us, in particular, to compare directly various QD systems.

*III.1.a Electrons*

The operator that accounts for the influence of confinement and strain and modify the conduction states by an energy $\Delta E_c$, reads $a|Z\rangle\langle Z|+b(|X\rangle\langle X|+|Y\rangle\langle Y|)$. After projection in the basis of irreducible representations and physical quantities of interest, this gives a series of energy shifts:

$$\Delta E_c = \sum_\alpha \sum_n a_\alpha^n \xi_\alpha^n (\Gamma_1) \ ,$$

where the summation on $\alpha=(k,s)$ extends over physical quantities of interest (here strain (*s*) and kinetic energy (*k*)), and where the summation over *n* extends over the number of different linear combinations of $\xi_\alpha^n$ terms that have the $\Gamma_1$ symmetry :

$$\Delta E_c = c_1 \varepsilon_{zz} + c_2 \left(\varepsilon_{xx}+\varepsilon_{yy}\right) + \hbar^2 \left( \frac{k_z^2}{2m_z^e} + \frac{k_x^2+k_y^2}{2m_{//}^e} \right) .$$

The parameters $c_1$ and $c_2$ are the conduction band deformation potentials (see Table 2).

*III.1.b Holes*

In the same spirit, the Hamiltonian that describes the valence band physics in the absence of spin writes in the most general way on the $\left(|p_x\rangle, |p_y\rangle, |p_z\rangle\right)$ basis:

$$\begin{pmatrix} \Xi_h + \Xi_1 + \Xi_{6a} & \Xi_{6b} & \Xi_{5a} \\ \Xi_{6b} & \Xi_h + \Xi_1 - \Xi_{6a} & \Xi_{5b} \\ \Xi_{5a} & \Xi_{5b} & \Xi_h - 2\Xi_1 \end{pmatrix} \quad (3)$$

The matrix elements and their symmetries are summarized in Table 1. Their expression and derivation are given in Appendix A.

It is important to outline that matrix element $\Xi_{6a}$ accounts, when any, for the deviation from in-plane isotropic shape via $\xi^a_{6k}=k_x^2-k_y^2$ and strain via $\xi^a_{6s}=\varepsilon_{xx}-\varepsilon_{yy}$. By properly choosing the x and y axes, one can disregard matrix element $\Xi_{6b}$ which may be set to zero via setting the symmetric combinations $\xi^b_{6\alpha}$ to zero.

After including the crystal field and the spin-orbit coupling (*see* Appendix A), the hole Hamiltonian writes in the $\Gamma_9$, $\Gamma_7^a$ and $\Gamma_7^b$ basis given in Table 3:

$$H_v = \Xi_h + \begin{pmatrix} \Xi_1+\Delta_1+\Delta_2 & -\Xi_{6a} & 0 & 0 & 0 & \frac{(-\Xi_{5a}-i\Xi_{5b})}{\sqrt{2}} \\ -\Xi_{6a} & \Xi_1+\Delta_1-\Delta_2 & \sqrt{2}\Delta_3 & 0 & 0 & \frac{(\Xi_{5a}-i\Xi_{5b})}{\sqrt{2}} \\ 0 & \sqrt{2}\Delta_3 & 0 & \frac{(\Xi_{5a}+i\Xi_{5b})}{\sqrt{2}} & \frac{(-\Xi_{5a}+i\Xi_{5b})}{\sqrt{2}} & 0 \\ 0 & 0 & \frac{(\Xi_{5a}-i\Xi_{5b})}{\sqrt{2}} & \Xi_1+\Delta_1+\Delta_2 & -\Xi_{6a} & 0 \\ 0 & 0 & \frac{(-\Xi_{5a}-i\Xi_{5b})}{\sqrt{2}} & -\Xi_{6a} & \Xi_1+\Delta_1-\Delta_2 & \sqrt{2}\Delta_3 \\ \frac{(-\Xi_{5a}+i\Xi_{5b})}{\sqrt{2}} & \frac{(\Xi_{5a}+i\Xi_{5b})}{\sqrt{2}} & 0 & 0 & \sqrt{2}\Delta_3 & 0 \end{pmatrix}$$

(4)

It is important to remark the main role of biaxial strain ($\Xi_1$) in GaN QDs, which decouples the $\Gamma_7^b$ state from the $\Gamma_9$ and $\Gamma_7^a$ states. Indeed the QDs are grown on top of a relaxed AlN buffer layer and a 2D GaN layer would present a 2.5 % compressive biaxial strain in the growth plane. The strain distribution in actual QDs is much more complex [35] but it is dominated by the same compressive strain. The term $\Xi_1$ is of the order of *200 meV* and is much larger than any other matrix element in the Hamiltonian $H_v$. These elements are roughly estimated in Appendix A for a typical QD corresponding to our experiments. It follows that the physics associated to the pair of $\Gamma_9$ and $\Gamma_7^a$ Bloch states can be treated independently of those originating from the $\Gamma_7^b$ states. The $\Gamma_7^b$ state is almost a pure p$_z$ state. The $\Gamma_9$ and $\Gamma_7^a$ Bloch states, lying at lower energy, are eigenstates of the following 2x2 matrix :

$\begin{pmatrix} \Xi_1+\Delta_1+\Delta_2 & \Xi_{6a} \\ \Xi_{6a} & \Xi_1+\Delta_1-\Delta_2 \end{pmatrix}$ which in turns is nothing but $\begin{pmatrix} \Delta_2 & \Xi_{6a} \\ \Xi_{6a} & -\Delta_2 \end{pmatrix}$.

The relevant parameter that rules the distribution of eigenstates between $p_x$ and $p_y$ states, i.e. the band mixing, is therefore the ratio $\frac{\Xi_{6a}}{\Delta_2}$. This ratio plays a major role in the interpretation of the polarization anisotropy of the optical response. The splitting between the two corresponding eigenstates is in first approximation equal to $2\sqrt{\Delta_2^2 + \Xi_{6a}^2}$.

**III.2 Few-carrier eigenstates**

*III.2.a) Spectral properties of neutral excitons*

The quantum state of the neutral exciton is a twelve-fold problem. The twelve exciton states are built from the electron and the hole states, and the many-body treatment of the problem leads to an effective Hamiltonian for the electron-hole spin-exchange interaction. In the most general way this writes [36] :

$$H_{exch} = +\frac{1}{2}\gamma \sigma_h . \sigma_e + \delta_z J_z^3 \sigma_z^e + \delta_{x+y}\left(J_x^3 \sigma_x^e + J_y^3 \sigma_y^e\right) + \delta_{x-y}\left(J_x^3 \sigma_x^e - J_y^3 \sigma_y^e\right) . \quad (5)$$

The first term is the spherical short-range exchange interaction [37]. The next three terms stand for the long-range exchange interaction [38,39]. The short-range term couples pairs of exciton states with hole and electron spins $\alpha\uparrow$ and $\beta\downarrow$, and identical orbital components. This term is proportional to the bulk value $\gamma=0.6 meV$ in GaN [40,41], which is more than 10 times larger than in bulk CdTe (*64 μeV*) or InAs (*0.29 μeV*) [42]. The long-range term implies more intricate selection rules. It is proportional to the longitudinal-transverse splitting $\Delta E_{LT} = 0.6 meV$ [43], i.e. to the oscillator strength [44], and it is also much larger than in CdTe or InAs.

The studies of InAs and CdTe QDs have shown over the past decade that the precise estimation of the exchange terms for excitons confined in a QD is complex. The experimental results cannot be compared quantitatively to theoretical models for each investigated QD with a given set of parameters, since the fine structure splitting depends on the QD conformation. These models are however useful to analyse statistical properties such as the distribution of splittings. This is why we do not choose the values of the 4 exchange parameters in Eq. 5 as fixed. The investigated QDs are very small ($L_z = 1.5 nm$), so that the overlap between electrons and holes is strong, as evidenced by their short radiative lifetime [15]. The exchange parameters are thus enhanced by the confinement [44-46,47]. Since bulk values are one order of magnitude stronger than in CdTe and InAs, we expect correspondingly larger values for the exchange parameters. For the following calculations, we use 1 meV for those parameters, for the sake of simplicity.

The derivation of the 12x12 matrix for the exchange interaction is presented in Appendix B. The full excitonic Hamiltonian $H_X = E_g + \Delta E_c - H_v + H_{exch}$ can be diagonalized into four blocks corresponding to the x, y, z polarized and dark exciton states, respectively, since the physics included in the present model respects the corresponding symmetries. The proper basis is presented in Table 4. The Hamiltonians for x and y-polarized states write :

$$H_{x,y} = E_g + \Delta E_c - \Xi_h - \begin{pmatrix} \Xi_1 + \Delta_1 + \Delta_2 & \pm \Xi_{6a} & 0 \\ \pm \Xi_{6a} & \Xi_1 + \Delta_1 - \Delta_2 & \sqrt{2}\Delta_3 \\ 0 & \sqrt{2}\Delta_3 & 0 \end{pmatrix}$$

$$+ \begin{pmatrix} \frac{\gamma}{2} - \frac{27}{16}\delta_z \pm \frac{3}{4}\delta_{x-y} & -\gamma + \frac{7}{8}\delta_{x+y} & \frac{7\sqrt{2}}{8}\delta_{x+y} \\ -\gamma + \frac{7}{8}\delta_{x+y} & \frac{\gamma}{2} + \frac{1}{16}\delta_z \pm \frac{3}{4}\delta_{x-y} & \pm \frac{7\sqrt{2}}{8}\delta_{x-y} \\ \frac{7\sqrt{2}}{8}\delta_{x+y} & \pm \frac{7\sqrt{2}}{8}\delta_{x-y} & -\frac{\gamma}{2} + \frac{1}{16}\delta_z \pm \frac{13}{8}\delta_{x-y} \end{pmatrix}$$

where − and + stand for x and y polarized states respectively. The matrices for the valence band and the exchange interaction are separated for clarity.

The Hamiltonian has been diagonalized for various sets of parameters in order to exhibit their respective influence. The oscillator strength of an eigenstate $|\psi\rangle$ simply writes in the basis of Table 4 as $\left|\frac{1}{\sqrt{2}}(\langle X_1|\psi\rangle + \langle X_2|\psi\rangle)\right|^2$, $\left|\frac{1}{\sqrt{2}}(\langle Y_1|\psi\rangle - \langle Y_2|\psi\rangle)\right|^2$, $\left|\langle Z_3|\psi\rangle\right|^2$ in the polarizations x, y and z, respectively. The results are summarized in the simulated absorption spectra (Fig. 4). PL spectra can be deduced by multiplying the absorption spectra with the Maxwell-Boltzmann occupation factor since we have shown that a thermal quasi-equilibrium is reached in each QD.

As explained in Section III.1.b., in our GaN QDs the term $\Xi_{1,s} \approx +200 meV$ dominates all other matrix elements. Hence in each polarization the third state is strongly split from the two first ones due to the large compressive strain, and we may restrict to the upper 2x2 sub-matrices in order to understand the main features.

The first observation is that the in-plane anisotropy term $\Xi_{6a}$ induces strong differences between the x and y oscillator strengths of excitons within each band. As qualitatively explained in Section III.1.b., the polarization degree of the first two transitions is governed by the ratio $\frac{\Xi_{6a}}{\Delta_2}$ of the in-plane anisotropy term to the valence band splitting, as demonstrated in Fig. 5a. It is only slightly affected when varying the exchange parameters. We notice that a rather small $\Xi_{6a}$ term of 10 meV is sufficient to induce a strong polarization of the transitions ($P=80\%$ as defined in Eq. 2).

The second observation is that the calculations yield a splitting between x and y-polarized excitons in the case of an anisotropic long-range exchange interaction $\delta_{x-y}$, as previously demonstrated for zinc-blende QDs. However, for GaN QDs, the splitting is also present if we simply include the short-range exchange interaction $\gamma$ and an in-plane anisotropy term $\Xi_{6a}$. This is a specificity of nitride QDs due to the proximity of the first two valence bands, since the short-range exchange interaction does not split the radiative doublet in zinc-blende QDs : it only contributes to the splitting between radiative and non-radiative states. In fact, in GaN QDs the ratio $\frac{\Xi_{6a}}{\Delta_2}$ also governs the value of the splitting, even if we only consider short-range interaction, as shown in Fig. 5b.

For the typical in-plane anisotropy deduced from the measured polarization degree, we calculate a value of the order of 1 meV for the short-range exchange term. In fact, the overall value of the fine structure splitting is roughly the sum of the short-range and anisotropic long-range contributions. Their relative signs is therefore crucial and it is *a priori* unknown [48]. We present two cases with same and opposite signs (Fig. 4.d,e), showing that the overall splitting may be quite large (3 meV) or small (less than 1 meV), compared to our spectral resolution. If nitride QDs present a statistical distribution of splittings as broad as the one that has been observed of InAs or CdTe QDs, then we can expect a mean value of a few meV in our system.

*III.2.b) Spectral properties of negatively charged trions*

Let us now consider the case of charged quantum dots. Here we focus on singly charged negative trions X⁻, but similar spectra would be obtained for positively charged trions X⁺.

The X⁻ trion consists in one hole and two $\Gamma_1$ electrons with opposite spins in the singlet state. Therefore the electron-hole exchange interaction vanishes. The symmetries of the trion states are those of the valence band, which simplifies significantly the calculation of allowed inter-band transitions. The final state after radiative recombination is preferentially a single electron in the ground state. The transitions in the emission spectrum are thus split as the valence band states are, and their oscillator strengths in the 3 polarizations are given by the $|p_x\rangle, |p_y\rangle, |p_z\rangle$ components of the valence band eigenstates.

The calculated absorption spectrum of a negatively charged QD is presented in Figs. 6.a,b in the absence or presence of an in-plane anisotropy term $\Xi_{6a}$. In the latter case only, the transitions are linearly polarized. The in-plane anisotropy term has the same effect as for the neutral exciton transitions. The ratio $\frac{\Xi_{6a}}{\Delta_2}$ governing the band mixing is still the relevant parameter. This polarization degree is reported in Fig. 7 as obtained from the diagonalization of the hole Hamiltonian $H_v$.

Let us mention that if the final state of the recombination consists in an electron in an excited confined state instead of the ground one, a new transition line may appear in the PL spectrum. The photons emitted via such Auger assisted transitions are subject to similar polarization properties.

Therefore, in the case of charged QDs, if a PL experiment is performed in temperature and excitation conditions that allow for simultaneous observation of such transitions, they will be observed in any polarization conditions without any polarization-related energy splitting, although intensity differences may be observed for different polarizations. The polarization degree of the transitions simply depends on the ratio $\frac{\Xi_{6a}}{\Delta_2}$.

**IV. COMPARISON BETWEEN EXPERIMENTAL AND THEORETICAL RESULTS**

The two main features of the polarization-resolved µPL results presented in Section II are the large values of polarization degree $P$ measured for each QD, and the absence of any doublet of purely linear and perpendicularly polarized transitions, within our experimental resolution of 1 meV.

Our model clearly shows that $P$ is determined by the ratio $\frac{\Xi_{6a}}{\Delta_2}$, whatever the nature of the emitting species – excitons or trions. The polarization degree presents a wide statistical distribution, which means that this ratio strongly varies from dot to dot in our sample. The strongest observed polarization degree (90 %) corresponds to an upper value of $\frac{\Xi_{6a}}{\Delta_2} \approx 2$, i.e. $\Xi_{6a} \approx 12\,meV$, and a ratio of 1, i.e. $\Xi_{6a} \approx 6\,meV$, appears as a typical value for the investigated QDs. The polarization direction of the PL reflects the proper axes for the anisotropy, which is also distributed and does not follow any specific crystallographic axis. We cannot distinguish between the two physical origins for this anisotropy : crystal deformation or QD shape. In the first case, the local deformation would not be perfectly biaxial in the growth plane. A uniaxial component of the deformation $\varepsilon_{xx} - \varepsilon_{yy} = 0.2\%$ is sufficient to induce a anisotropic term $\Xi_{6a,s} \approx 6\,meV$ (to be compared to the total in-plane deformation of 2.5% of GaN on AlN). Assuming such a feature is reasonable and supported by the observation of inhomogeneous deformation at the micrometer scale through Raman studies [49] and cathodoluminescence imaging [50,51], as well as the modelling of the inhomogeneous deformation at the nanometer scale within a GaN QD [35,52]. In the case of shape anisotropy, the QDs would be elongated along an arbitrarily oriented axis. A more precise **k.p** modelling of the confinement in (In,Ga)N QDs recently predicted that a shape anisotropy $(L_x - L_y)/L_x$ of only 10 % is sufficient to explain a polarization degree of 50 % [26].

We may emphasize that similar anisotropy terms of the order of 10 meV would lead to a very small polarization degree in InAs or CdTe QDs since the splitting $\Delta_{HH-LH}$ between heavy-hole and light-hole confined exciton states is much greater than the A-B splitting $2\Delta_2$ in GaN, and therefore the ratio $\Xi_{6a}/\Delta_{HH-LH}$ is only worth a few percents in those material systems (*see* Fig. 5). This is due to the strong difference between heavy- and light-hole effective masses in these materials. Large polarization degree has only been observed in strongly elongated InAs QDs [53], in CdTe QDs grown on vicinal substrates [54] or in quantum wires [55], that is to say when the confinement is highly anisotropic and mixes valence bands. In nitride-based heterostructures, A and B valence bands have identical on-axis effective masses, *i.e.* identical vertical confinement, and differ only by the in-plane masses, inducing small differences in the lateral confinement.

The random direction of the polarization may be related to the existence of 3 couples of equivalent crystallographic axes in wurtzite materials, instead of only one in zinc-blende materials. However only two preferential axes have been identified for (In,Ga)N QDs [26] due to the strain induced by the buffer layer, in contrast with our results. This evidences that the polarization properties of the emission reflects the structure and the strain of the layers lying under the quantum dots. We could also remark here that single-dot linear polarization along non-crystallographic axes has recently been reported by E. Poem *et al.* [7, 56] for excitonic complexes $X^{-2}$, $X_2^-$, $X_2^+$; the authors tentatively ascribe their failure in describing this phenomenon to the one-band model that they developed.

Concerning the absence of experimentally-resolved doublets, our model suggests that either the fine-structure splitting is smaller than 1 meV, or we only observe trions by µPL on this sample. According to our model, fine structure splittings smaller than 1 meV can be understood only if we assume very specific combinations and signs of the different contributions to the spin-exchange interaction. Therefore, assuming that QD emission arises from neutral excitons would imply that such a situation is realized in all the 20 measured QDs. Alternatively, we may observe various charge states of the QDs, so that some of the investigated transitions would belong to initially neutral QDs, and others to charged QDs. Further works will be required in order to independently measure the charge state of the unexcited QDs, and to develop a quantitative model of the fine structure in nitride QDs.

Our observations have important consequences on the potential applications of nitride QDs as controlled single photon emitters. For example, our findings seem to disqualify this system for the development of semiconductor sources of triggered entangled photon pairs [57-59]. Indeed, for the latter, unpolarized emission is crucial, unless a very efficient control of the anisotropy of each QD is accessible. If one aims at achieving such a control, our results indicate that it will have to be much more accurate than for InAs QDs [60-62]. This is due to the small splitting between A and B valence bands in GaN. Nevertheless, this feature does not preclude the realization of single photon sources based on GaN QDs [17].

**V. CONCLUSION**

We have observed a strong linear polarization degree in the PL spectra of single GaN/AlN QDs, which reflects the complexity of valence band mixing. The latter is due to an in-plane anisotropy of the shape and/or the deformation field of each QD. The key parameter that controls the magnitude of this effect is the ratio between the matrix element accounting for this anisotropy, and the splitting between the first two valence bands, allowing for an interesting comparison with zinc-blende QDs studied so far.

We interpret the absence of any energy splitting between cross-polarized transitions in the measured PL spectra, within the spectral resolution (1 meV), as a possible evidence of either trion emission, *i.e.* of charged QDs, or of a specific combination of the short- and long-range exchange terms leading to a small fine structure splitting for neutral excitons.

We base our reasoning on a model of the exciton and trion fine structures. The three valence bands, the spin-exchange interaction (for neutral excitons), and the symmetry breaking confinement and strain terms are included in a simple matrix Hamiltonian. This allows for a comparison between all QD systems. The model evidences that neutral exciton spectra in GaN QDs are very different from those of InAs or CdTe QDs: the exchange splitting is, indeed, due to anisotropic long-range spin-exchange interaction, but it also results from the combination of a weak in-plane anisotropy and short-range exchange interaction, which is specific to GaN QDs due to their strong band mixing between A and B valence bands.

**ACKNOWLEDGMENTS**

This work was supported by the French Ministry of Education, Research and Technology within the "BOQUANI" and "BUGATI" Research Programs.

# Appendix A : Valence band Hamiltonian

Let us consider the Hamiltonian for holes in the valence band and give the expression of each matrix element in Eq. (3) individually as a function of the strain and confinement variables presented in Table 1.

*A.1) Estimation of confinement-related terms*

As presented in the introduction of the model, the confinement is deliberately accounted for through a very rough approximation : we consider a typical quantum dot of height $L_z=1.5\,nm$ and an aspect ratio $L_z/L_{x,y} \approx 0.2$ [63]. Its lateral size is therefore typically $L_x \approx L_y \approx 7\,nm$ if we temporarily exclude in-plane anisotropy. The wave-vector components in the expression of the kinetic energy terms are simply substituted by $k_{x,y,z}=\pi/L_{x,y,z}$ for the fundamental state, giving the correct order of magnitude. This approximation is justified since the knowledge of the precise spatial wavefunction and of the confinement energy is not crucial in our model, and we focus on the valence band mixing and the symmetry of the eigenstates in the ground confined mode.

Moreover we assume equal masses for holes in the 3 states $\left(|p_x\rangle, |p_y\rangle, |p_z\rangle\right)$. This is rigorous for on-axis masses $m_z$ but it is false for in-plane masses ($m_{//}$ along *x,y*) (see Table 2). However the vertical confinement is much stronger than the lateral one due to the small aspect ratio of the QD shape. The mass $m_c$ involved in cross-coupling terms is fairly unknown up to now, and is assumed to be of the order of $m_{//}$ [64].

The kinetic energy terms are therefore the following : the average confinement energy is $\Xi_{h,k} = \dfrac{\hbar^2}{2m_{//}}\left(k_x^2 + k_y^2\right) + \dfrac{\hbar^2}{2m_z}k_z^2$; the cross-coupling terms, accounting for the confinement-induced band-mixing for any QD shape, are $\Xi_{5a,k} = \dfrac{\hbar^2}{2m_c}k_x k_z$, $\Xi_{5b,k} = \dfrac{\hbar^2}{2m_c}k_y k_z$, $\Xi_{6b,k} = \dfrac{\hbar^2}{2m_c}k_x k_y$; the anisotropy of the in-plane confinement ($L_x \neq L_y$) is responsible for the last term $\Xi_{6a,k} = \dfrac{\hbar^2}{2m_{//}}\left(k_x^2 - k_y^2\right)$.

The order of magnitude of the off-diagonal terms is the following :
$\Xi_{5a,k} \approx \Xi_{5b,k} \approx 36\,meV$, $\Xi_{6b,k} \approx 8\,meV$, $\Xi_{6a,k} \leq 8\,meV$.

$\Xi_{6b,k}$ can be set to zero if choosing an appropriate set of the x and y axes : indeed the bilinear form corresponding to $\Gamma_6$ kinetic energy terms writes :

$$\Xi_6 = b_{6a}\left(k_x^2 - k_y^2\right) + 2b_{6a}k_x k_y$$
$$= \left(b_{6a}\cos 2\theta + b_{6b}\sin 2\theta\right)\left(k_{x'}^2 - k_{y'}^2\right) + 2\left(-b_{6a}\sin 2\theta + b_{6b}\cos 2\theta\right)k_{x'}k_{y'},$$

where *x'* and *y'* axis are deduced from *x* and *y* axis by a rotation of angle $\theta$ around the z axis. Choosing $\theta = \arctan(b_{6b}/b_{6a})/2$ allows to set the second term to zero. The same procedure can be performed if we include simultaneously kinetic and strain terms, and the polarization angle $\theta$ and $\theta + \pi/2$ of the hole states depends on the combination of all $\Gamma_6$ kinetic energy and strain terms.

*A.2) Estimation of strain-related terms*

The strain-related matrix elements are obtained from the deformation tensor and the valence band deformation potentials given in Table 2 :

$$\Xi_{h,s} = D_1\varepsilon_{zz} + D_2(\varepsilon_{xx} + \varepsilon_{yy}) + 2/3\left(C_3\varepsilon_{zz} + C_4(\varepsilon_{xx} + \varepsilon_{yy})\right);$$

$$\Xi_{1,s} = 1/3\left(C_3\varepsilon_{zz} + C_4(\varepsilon_{xx} + \varepsilon_{yy})\right);$$

$$\Xi_{6a,s} = -C_5(\varepsilon_{xx} - \varepsilon_{yy}).$$

Within the quasi-cubic approximation, in-plane and out-of-plane deformations are related through the elastic coefficients :

$$\varepsilon_{zz} = -C_{13}/C_{33}(\varepsilon_{xx} + \varepsilon_{yy}).$$

The samples being grown along the (0001) axis, the shear deformation is assumed to vanish : $\Xi_{5a,s} = \Xi_{5b,s} = \Xi_{6b,s} = 0$.

Assuming that GaN experiences a strong biaxial compression on a relaxed AlN buffer layer and mainly presents an in-plane biaxial deformation of $\varepsilon_{xx} \approx \varepsilon_{yy} \approx -2.5\%$, the order of magnitude of the strain-related terms is thus the following :

$\Xi_{h,s} \approx -430\ meV$, $\Xi_{1,s} \approx +200\ meV$. (Here we include the contribution of the corresponding term for electrons, since only the sum of electron and hole deformation potentials is known).

As explained in the text, if we now consider the hole spin and the basis given in Table 4, the off-diagonal matrix elements (see Eq. (4)) which couple the $\Gamma_7^b$ state to the $\Gamma_9$ and $\Gamma_7^a$ states are therefore much smaller than the on-diagonal splitting $\Xi_{1,s}$. The band mixing of the $\Gamma_7^b$ state with the two other ones is negligible and a model restricted to the $\Gamma_9$ and $\Gamma_7^a$ states provides in first approximation a simple and correct description of the PL spectra of GaN/AlN QDs. However all the spectra presented in this work are calculated for the complete set of valence bands.

Finally the value of the anisotropic deformation term is for example $\Xi_{6a,s} \approx -3 meV$ for a small in-plane uniaxial deformation $(\varepsilon_{xx} - \varepsilon_{yy}) = 0.1\%$.

*A.3) Full Hamiltonian for holes*

Starting from Eq. 3, we have to include the crystal field coupling which writes $\Delta_1 L_z^2$. The spin-less valence band Hamiltonian becomes in the $(|p_x\rangle, |p_y\rangle, |p_z\rangle)$ basis:

$$\begin{pmatrix} \Xi_h + \Xi_1 + \Delta_1 + \Xi_{6a} & 0 & \Xi_{5a} \\ 0 & \Xi_h + \Xi_1 + \Delta_1 - \Xi_{6a} & \Xi_{5b} \\ \Xi_{5a} & \Xi_{5b} & \Xi_h - 2\Xi_1 \end{pmatrix}$$

Next, we have to account for the existence of spin in the building-up of the appropriate basis, so as to include the two-component spin-orbit interaction $H_{so} = \Delta_2 \left( L_x \sigma_x^h + L_y \sigma_y^h \right) + \Delta_3 L_z \sigma_z^h$. In the $\Gamma_9$, $\Gamma_7^a$ and $\Gamma_7^b$ basis given in Table 3, we obtain the full Hamiltonian for holes presented in Eq. 4 (*see* text).

*A.4) Role of spontaneous and piezoelectric polarization fields*

The polarization fields induce a large electric field in GaN nanostructures and have a strong impact on the confined states. The spontaneous polarization field of GaN is collinear to the growth direction in our samples. The components of the piezoelectrically-induced polarization **P** are given in the (**x,y,z**) basis by:

$$\begin{bmatrix} P_x \\ P_y \\ P_z \end{bmatrix} = \begin{pmatrix} 0 & 0 & 0 & 0 & e_{15} & 0 \\ 0 & 0 & 0 & e_{15} & 0 & 0 \\ e_{13} & e_{13} & e_{33} & 0 & 0 & 0 \end{pmatrix} \begin{bmatrix} \varepsilon_{xx} \\ \varepsilon_{yy} \\ \varepsilon_{zz} \\ \varepsilon_{yz} \\ \varepsilon_{xz} \\ \varepsilon_{xy} \end{bmatrix}.$$

This equation indicates that anisotropy of in-plane strain does not impact the x and y components of the piezo-electric polarization. Therefore the total polarization field is along the z axis. The corresponding matrix element has a $\Gamma_1$ symmetry and contributes only to $\Xi_h$ through the vertical confinement term in the Hamiltonian for holes.

# Appendix B : Spin-exchange Hamiltonian

*B.1) Short-range exchange interaction*

The short-range exchange interaction has to be written in the exciton basis built from Table 3, and reduces in the spin basis $(\alpha\uparrow,\beta\downarrow,\alpha\downarrow,\beta\uparrow)$ to :

$$H_{exch,SR} = +\frac{1}{2}\gamma\vec{\sigma}_e.\vec{\sigma}_h = \begin{pmatrix} \gamma/2 & 0 & 0 & 0 \\ 0 & \gamma/2 & 0 & 0 \\ 0 & 0 & -\gamma/2 & \gamma \\ 0 & 0 & \gamma & -\gamma/2 \end{pmatrix}.$$

*B.2) Long-range exchange interaction*

From Eq. (5), the long-range exchange interaction writes in the $|J_h, m_{Jh}, m_e\rangle$ basis

$$H_{exch,LR} = \begin{pmatrix} H_{LR,+} & 0 \\ 0 & H_{LR,-} \end{pmatrix}, \text{ with}$$

$$H_{LR,\pm} = \begin{pmatrix} \pm\frac{27}{16}\delta_z & \frac{7}{8}\sqrt{3}\delta_{x\mp y} & 0 & \frac{3}{4}\delta_{x\pm y} & 0 & 0 \\ \frac{7}{8}\sqrt{3}\delta_{x\mp y} & \frac{1}{16}\delta_z & 0 & 0 & \frac{5}{2}\delta_{x\pm y} & 0 \\ 0 & 0 & \frac{1}{16}\delta_z & 0 & 0 & \frac{1}{8}\delta_{x\pm y} \\ \frac{3}{4}\delta_{x\pm y} & 0 & 0 & \pm\frac{27}{16}\delta_z & \frac{7}{8}\sqrt{3}\delta_{x\mp y} & 0 \\ 0 & \frac{5}{2}\delta_{x\pm y} & 0 & \frac{7}{8}\sqrt{3}\delta_{x\mp y} & \frac{1}{16}\delta_z & 0 \\ 0 & 0 & \frac{1}{8}\delta_{x\pm y} & 0 & 0 & \frac{1}{16}\delta_z \end{pmatrix}.$$

The basis is ordered as follows :

$$\left(\left|\frac{3}{2},\frac{3}{2},\uparrow\right\rangle, \left|\frac{3}{2},\frac{1}{2},\downarrow\right\rangle, \left|\frac{1}{2},\frac{1}{2},\downarrow\right\rangle, \left|\frac{3}{2},-\frac{3}{2},\downarrow\right\rangle, \left|\frac{3}{2},-\frac{1}{2},\uparrow\right\rangle, \left|\frac{1}{2},-\frac{1}{2},\uparrow\right\rangle, \left|\frac{3}{2},\frac{3}{2},\downarrow\right\rangle, \left|\frac{3}{2},\frac{1}{2},\uparrow\right\rangle, \left|\frac{1}{2},\frac{1}{2},\uparrow\right\rangle, \left|\frac{3}{2},-\frac{3}{2},\uparrow\right\rangle, \left|\frac{3}{2},-\frac{1}{2},\downarrow\right\rangle, \left|\frac{1}{2},-\frac{1}{2},\downarrow\right\rangle\right)$$

It has to be noticed that this interaction leaves the 6 z-polarized and dark states, and the 6 x and y-polarized states, uncoupled. Moreover, it is compatible with the 3x3 block-diagonalization suggested by the valence-band Hamiltonian (*see* text).

*B.3) Full Hamiltonian for excitons*

The full excitonic Hamiltonian $H_X = E_g + \Delta E_c - H_v + H_{exch}$ can be diagonalized into four blocks corresponding to the x, y, z polarized and dark exciton states by using the proper basis presented in Table 4. The Hamiltonians for z-polarized and dark excitons are similar and write :

$$H_{z,\varnothing} = E_g + \Delta E_c - \Xi_h - \begin{pmatrix} \Xi_1 + \Delta_1 + \Delta_2 & \Xi_{6a} & 0 \\ \Xi_{6a} & \Xi_1 + \Delta_1 - \Delta_2 & \sqrt{2}\Delta_3 \\ 0 & \sqrt{2}\Delta_3 & 0 \end{pmatrix}$$

$$+ \begin{pmatrix} -\dfrac{\gamma}{2} + \dfrac{27}{16}\delta_z \pm \dfrac{3}{4}\delta_{x+y} & \gamma \pm \dfrac{7}{8}\delta_{x-y} & \dfrac{7\sqrt{2}}{8}\delta_{x-y} \\ \gamma \pm \dfrac{7}{8}\delta_{x-y} & -\dfrac{\gamma}{2} - \dfrac{1}{16}\delta_z \pm \dfrac{3}{4}\delta_{x+y} & \dfrac{7\sqrt{2}}{8}\delta_{x+y} \\ \dfrac{7\sqrt{2}}{8}\delta_{x-y} & \dfrac{7\sqrt{2}}{8}\delta_{x+y} & (1\mp 2)\dfrac{\gamma}{2} - \dfrac{1}{16}\delta_z \pm \dfrac{13}{8}\delta_{x+y} \end{pmatrix}$$

where – and + stand for z-polarized and dark states respectively. The one for x and y polarized are similar, and are given in the main text.

at the same energy as our QDs. The enhancement of the anisotropic long-range exchange interaction $\delta_{x\text{-}y}$ cannot be simply estimated without a precise knowledge of the lateral confinement of excitons in the QD.

# TABLES

| Symmetry | Variables | | | Matrix element | |
|---|---|---|---|---|---|
| | Strain | Kinetic energy | Notation | For electrons | For holes |
| $\Gamma_1$ | $\varepsilon_{zz}$ : biaxial deformation | $k_z^2$ : Vertical confinement | $\xi_1^a$ | $\Delta E_c$ | $\Xi_h + \Xi_1$ for $p_x, p_y$ holes, |
| | $\varepsilon_{xx} + \varepsilon_{yy}$ : biaxial deformation | $k_x^2 + k_y^2$ : Average in-plane confinement | $\xi_1^b$ | | $\Xi_h - 2\Xi_1$ for $p_z$ holes, |
| $\Gamma_5$ | $\varepsilon_{xz}, \varepsilon_{yz}$ : shear deformation, zero in our case | $k_x k_z$, $k_y k_z$ : Cross-coupling terms | $\xi_5^a, \xi_5^b$ | | $\Xi_{5a}, \Xi_{5b}$ |
| $\Gamma_6$ | $\varepsilon_{xx} - \varepsilon_{yy}$ : uniaxial deformation | $k_x^2 - k_y^2$ : in-plane shape anisotropy | $\xi_6^a$ | | $\Xi_{6a}$ |
| | $\varepsilon_{xy}$ : shear deformation, zero in our case | $2 k_x k_y$ : Cross-coupling terms | $\xi_6^b$ | | $\Xi_{6b} = 0$ |

Table 1 : Typical symmetrized combinations of the strain tensor, of the bilinear function of components of the wave vector, and a generalized notation of them; corresponding matrix elements in the hamiltonian. The dominant terms are highlighted.

| Parameter | Value | Reference |
|---|---|---|
| GaN Bandgap | $E_g = 3.48\,eV$ | [64] |
| Conduction band effective mass | $m_{e,z} = m_{e,//} = 0.2\,m_0$ | [64] |
| Valence band parameters | $\Delta_1 = 8.7\,meV$ <br> $\Delta_2 = \Delta_3 = 5.7\,meV$ | [40] |
| Valence band effective masses | $m_{h,z} = m_{h,//} = m_{h,c} \approx m_0$ | [64] |
| Deformation potentials | $C_1 + D_1 = -5.32\,eV$ <br> $C_2 + D_2 = -10.23\,eV$ <br> $D_3 = -2D_4 = -4.91\,eV$ | [65] |
| | $D_5 \approx -3\,eV$ | [66-68] |
| Elastic constants | $C_{13} = 114\,GPa$ <br> $C_{33} = 381\,GPa$ | [69] |

Table 2 : Material parameters of bulk GaN used in the model.

| | | |
|---|---|---|
| $\Gamma_9$ | $Y_1^1\alpha = -(p_x+ip_y)/\sqrt{2}\,\alpha = \left|\frac{3}{2},\frac{3}{2}\right\rangle$ | $Y_1^{-1}\beta = (p_x-ip_y)/\sqrt{2}\,\beta = \left|\frac{3}{2},-\frac{3}{2}\right\rangle$ |
| $\Gamma_7^a$ | $Y_1^{-1}\alpha = (p_x-ip_y)/\sqrt{2}\,\alpha = \sqrt{\frac{1}{3}}\left|\frac{3}{2},-\frac{1}{2}\right\rangle - \sqrt{\frac{2}{3}}\left|\frac{1}{2},-\frac{1}{2}\right\rangle$ | $Y_1^1\beta = -(p_x+ip_y)/\sqrt{2}\,\beta = \sqrt{\frac{1}{3}}\left|\frac{3}{2},\frac{1}{2}\right\rangle + \sqrt{\frac{2}{3}}\left|\frac{1}{2},\frac{1}{2}\right\rangle$ |
| $\Gamma_7^b$ | $Y_1^0\beta = p_z\beta = \sqrt{\frac{2}{3}}\left|\frac{3}{2},-\frac{1}{2}\right\rangle + \sqrt{\frac{1}{3}}\left|\frac{1}{2},-\frac{1}{2}\right\rangle$ | $Y_1^0\alpha = p_z\alpha = \sqrt{\frac{2}{3}}\left|\frac{3}{2},\frac{1}{2}\right\rangle - \sqrt{\frac{1}{3}}\left|\frac{1}{2},\frac{1}{2}\right\rangle$ |

Table 3 : Expression giving the symmetrized combinations of wave functions (p states and spherical harmonics) giving the valence band wave functions in the context of the double group representation, as well as the correspondence with the $\left|J_h, m_{Jh}\right\rangle$ eigenstates of $J_h = L + \sigma_h$. The spin components of the hole are $\sigma_z^h = \alpha, \beta$.

| State | Composition | Symmetry |
|---|---|---|
| $|Z_1\rangle$ | $\frac{1}{\sqrt{2}}(Y_1^1\alpha\uparrow + Y_1^{-1}\beta\downarrow)$ | $\Gamma_6$ |
| $|Z_2\rangle$ | $\frac{1}{\sqrt{2}}(Y_1^{-1}\alpha\uparrow + Y_1^1\beta\downarrow)$ | $\Gamma_1$ |
| $|Z_3\rangle$ | $\frac{1}{\sqrt{2}}(Y_1^0\alpha\downarrow + Y_1^0\beta\uparrow)$ | $\Gamma_1$ |
| $|\varnothing_1\rangle$ | $\frac{1}{\sqrt{2}}(Y_1^1\alpha\uparrow - Y_1^{-1}\beta\downarrow)$ | $\Gamma_6$ |
| $|\varnothing_2\rangle$ | $\frac{1}{\sqrt{2}}(Y_1^{-1}\alpha\uparrow - Y_1^1\beta\downarrow)$ | $\Gamma_2$ |
| $|\varnothing_3\rangle$ | $\frac{1}{\sqrt{2}}(Y_1^0\alpha\downarrow - Y_1^0\beta\uparrow)$ | $\Gamma_2$ |
| $|X_1\rangle$ | $\frac{1}{\sqrt{2}}(Y_1^1\alpha\downarrow + Y_1^{-1}\beta\uparrow)$ | $\Gamma_5$ |
| $|X_2\rangle$ | $\frac{1}{\sqrt{2}}(Y_1^{-1}\alpha\downarrow + Y_1^1\beta\uparrow)$ | $\Gamma_5$ |
| $|X_3\rangle$ | $\frac{1}{\sqrt{2}}(Y_1^0\alpha\uparrow + Y_1^0\beta\downarrow)$ | $\Gamma_5$ |
| $|Y_1\rangle$ | $\frac{1}{\sqrt{2}}(Y_1^1\alpha\downarrow - Y_1^{-1}\beta\uparrow)$ | $\Gamma_5$ |
| $|Y_2\rangle$ | $\frac{1}{\sqrt{2}}(Y_1^{-1}\alpha\downarrow - Y_1^1\beta\uparrow)$ | $\Gamma_5$ |
| $|Y_3\rangle$ | $\frac{1}{\sqrt{2}}(Y_1^0\alpha\uparrow - Y_1^0\beta\downarrow)$ | $\Gamma_5$ |

Table 4 : Excitonic wave functions built from the valence hole ones and the electron one. Arrows are electron spin up and down eigenvectors, while $\alpha$ and $\beta$ are the hole spin components. The groups of 3 consecutive states form the basis in which the Hamltonians $H_{z,\varnothing}$ and $H_{x,y}$ are written in section III.2.a.

# FIGURES

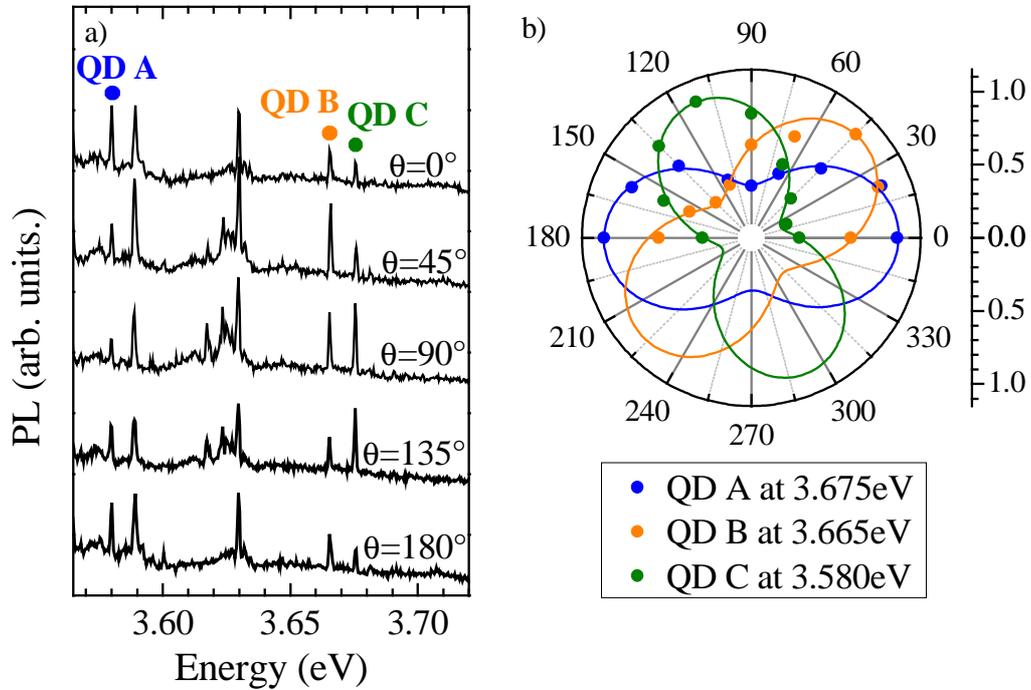

Figure 1 : (a) µPL spectra of a few QDs, as a function of the angle of the polarizer analyzing the PL. The spectra are vertically offset for clarity. (b) Polar representation of the normalized intensity of the 3 lines labeled A, B and C (circles), and their fit by equation 1 (line).

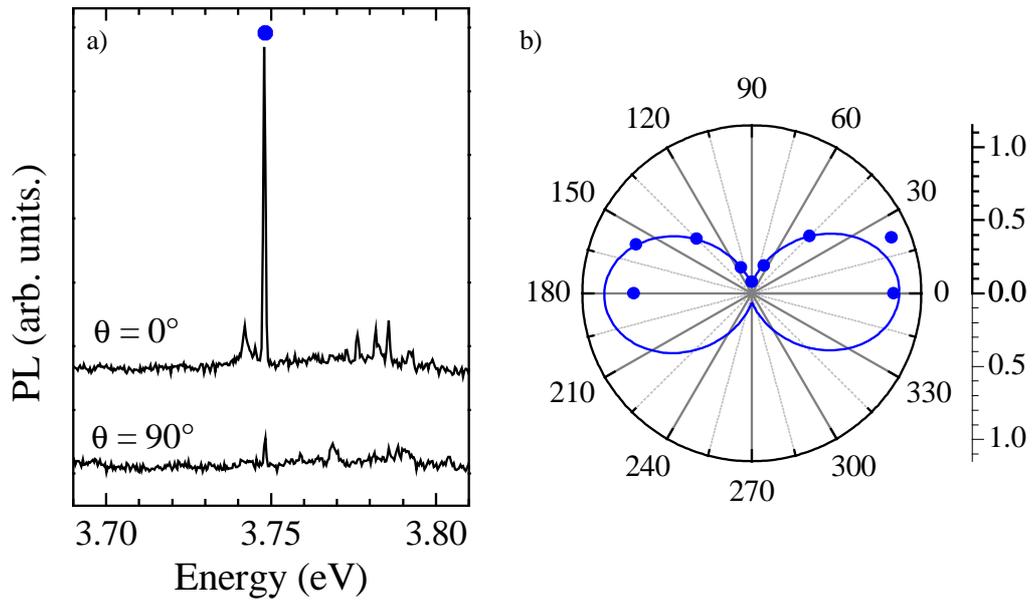

Figure 2 : (a) µPL spectra of a single quantum dot for two perpendicular polarizations. The spectra are vertically offset for clarity. (b) Polar representation of the normalized intensity of the main line (circles), and its fit by equation 1 (line). The polarization degree is $P=90\%$.

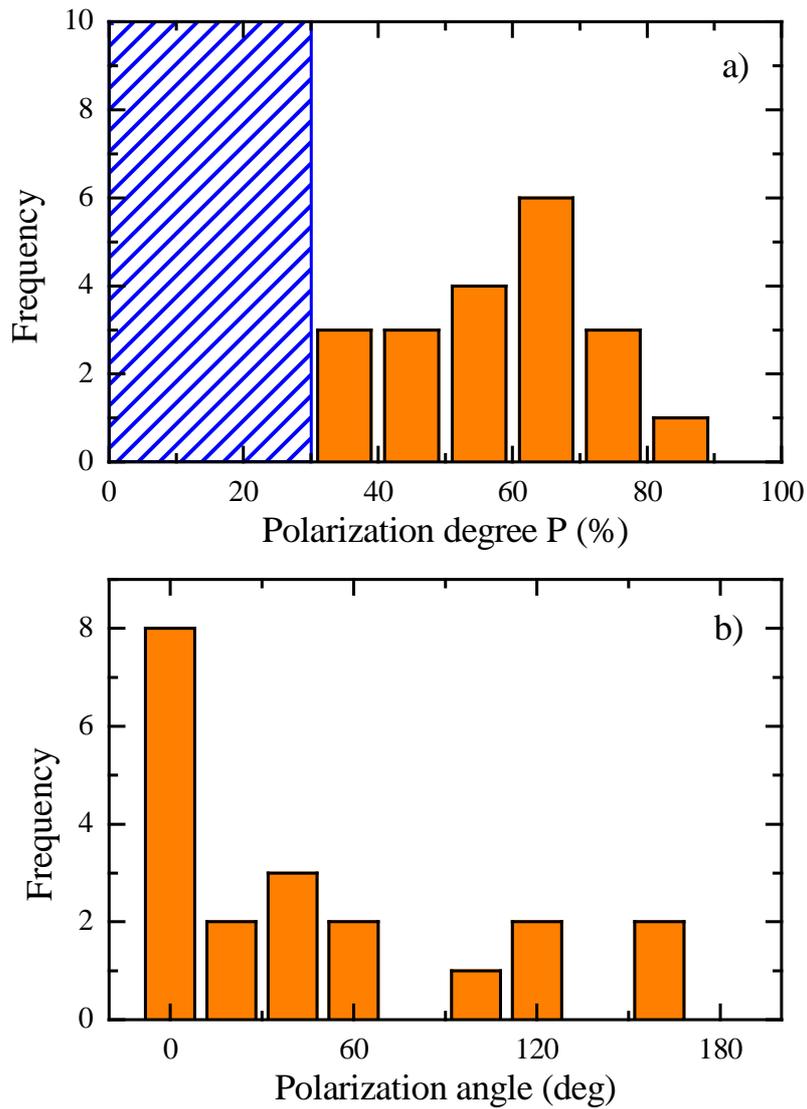

Figure 3 : a) Statistics of the polarization degree of the PL of 20 QDs recorded at 8 different positions of the same sample. The dashed area illustrates that polarization degrees smaller than 30% could not be unambiguously determined from the experimental results. b) Statistics of the measured polarization angles.

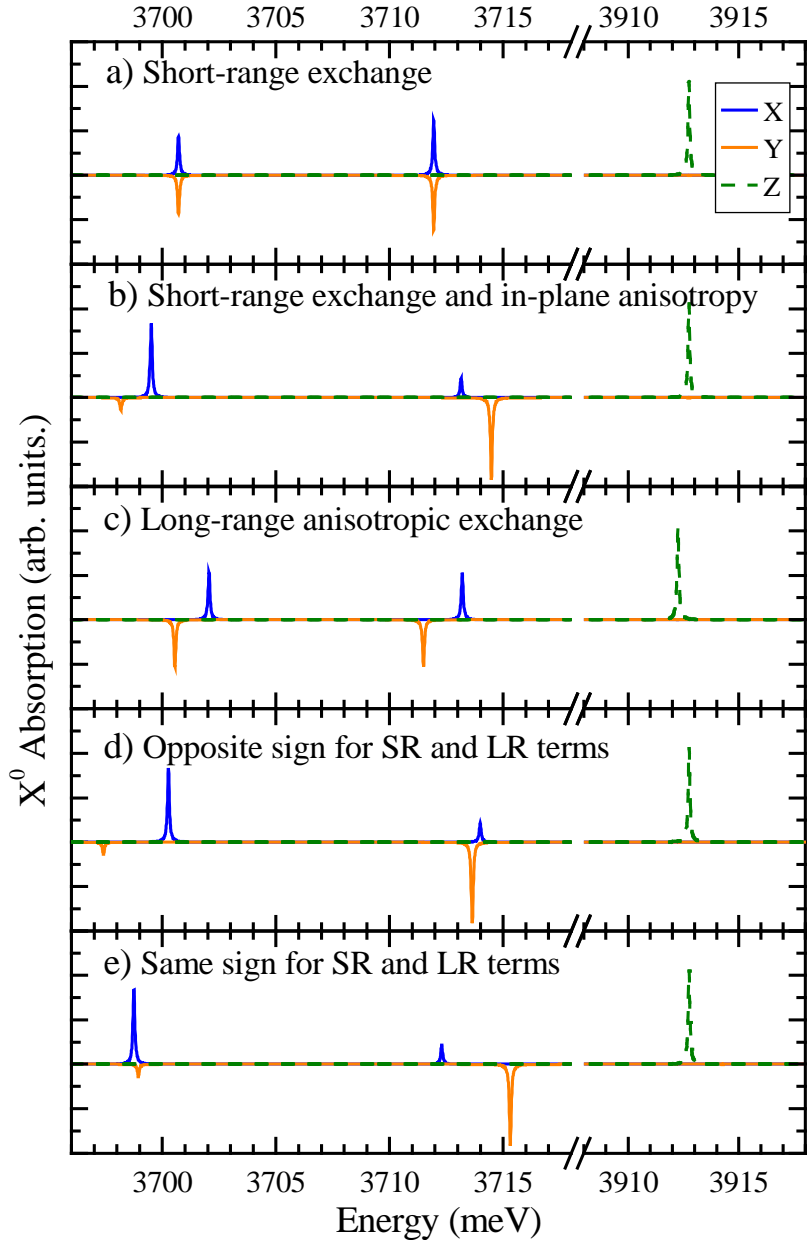

Figure 4 : Calculated neutral exciton absorption spectrum of a initially neutral quantum dot. a) with only the short-range exchange interaction $\gamma = 1\,meV$ ; b) with both short-range exchange interaction and an in-plane anisotropy term $\Xi_{6a} = 5\,meV$ ; c) with only the long-range exchange interaction term $\delta_{x-y} = 1\,meV$ ; d) and e) with all terms, choosing short-range long-range terms with opposite or identical signs (*see* text). The spectra are calculated with a homogeneous broadening of $0.1\,meV$. The spectra in the polarization y are inverted for clarity.

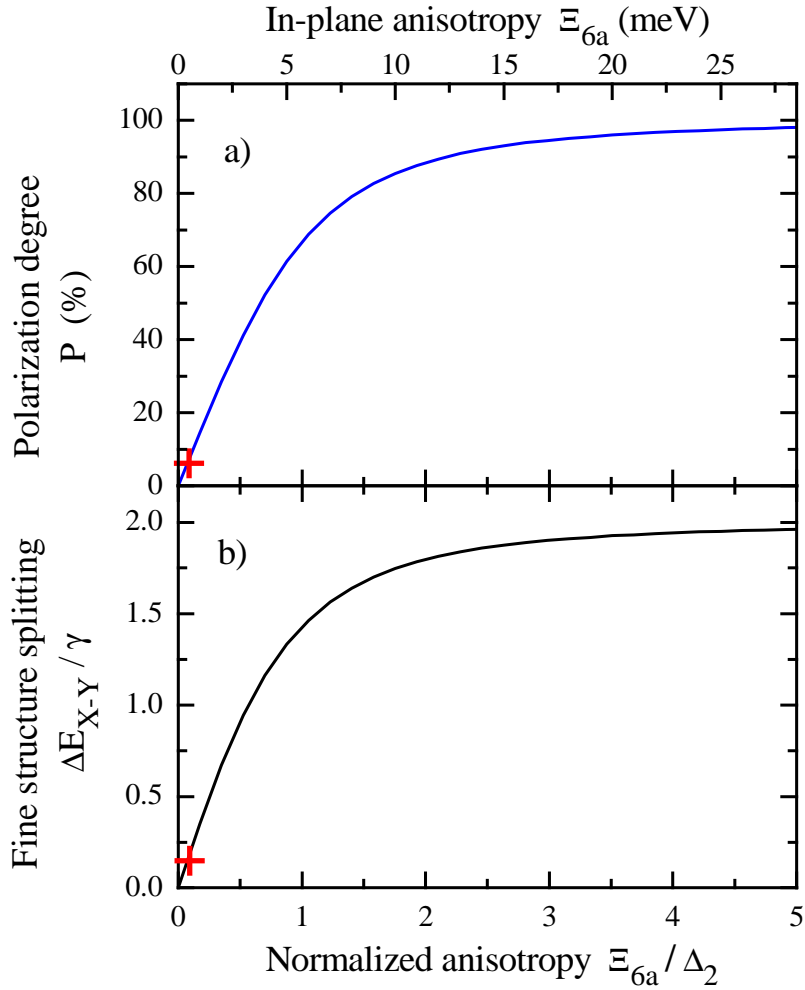

Figure 5 : Role of the normalized in-plane anisotropy $\frac{\Xi_{6a}}{\Delta_2}$ determining the band mixing, on the polarization anisotropy (a) and on the contribution of the short-range exchange interaction to the fine structure splitting (b) of the ground excitonic states. The cross (+) near the origin of each panel illustrates the role of a similar in-plane anisotropy term $\Xi_{6a} = 10\,meV$ on the properties of a "typical" InAs QD which confinement would induce a heavy hole-light hole splitting of $\Delta_{HH-LH} = 100\,meV$ : the degree of linear polarization is weak and the short-range exchange interaction does not significantly contribute to the fine structure splitting.

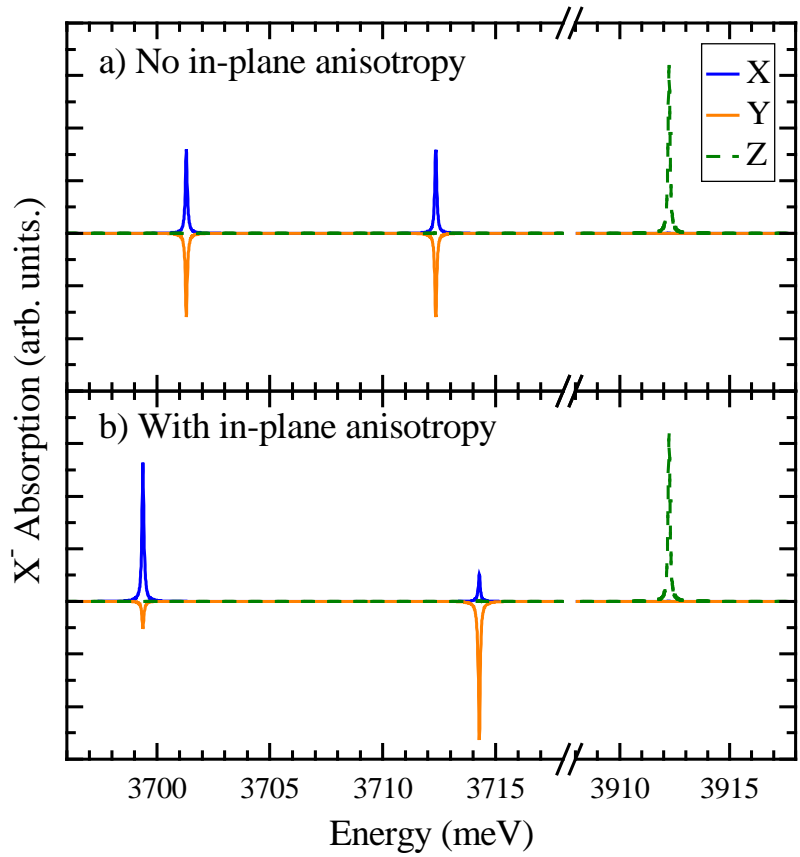

Figure 6 : Trion absorption spectrum of an initially negatively charged quantum dot a) in the absence of in-plane anisotropy; b) for an in-plane anisotropy term $\Xi_{6a} = 5\,meV$. The spectra are calculated with a homogeneous broadening of $0.1\,meV$. The spectra in the polarization y are inverted for clarity.

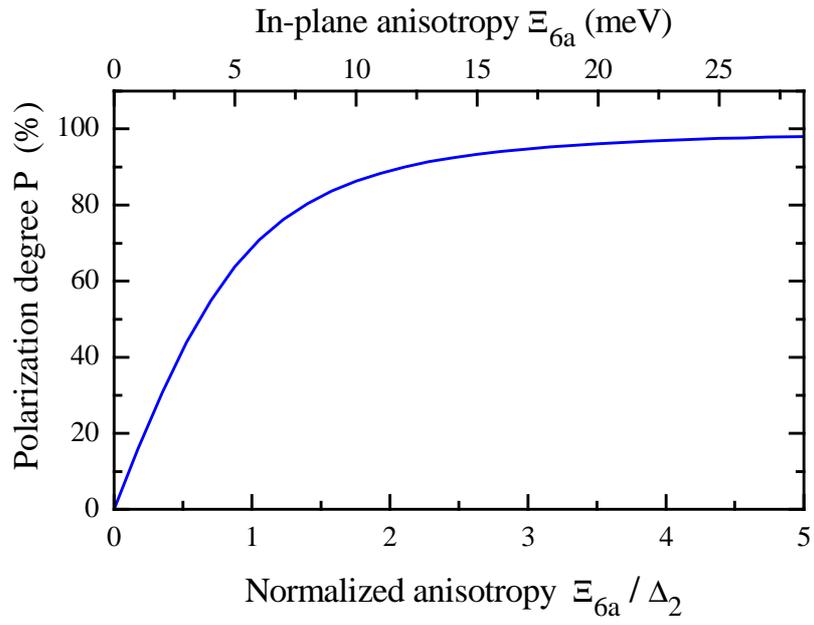

Figure 7 : Polarization degree of the first transition of the negative trion absorption spectrum, as a function of the normalized in-plane anisotropy $\frac{\Xi_{6a}}{\Delta_2}$ of the shape and/or strain of the QD.